# The Arabic Citation Index - Toward a better understanding of Arab scientific literature

Jamal El-Ouahi[1,2]

[1] j.el.ouahi@cwts.leidenuniv.nl - https://orcid.org/0000-0002-3458-7503
Centre for Science and Technology Studies (CWTS), Leiden University, Leiden, Netherlands

[2] Clarivate Analytics, Dubai Internet City, Dubai, United Arab Emirates

## Abstract

The Arabic Citation Index (ARCI) was launched in 2020. This article provides an overview of the scientific literature contained in this new database and explores its possible usage in research evaluation. As of May 2022, ARCI had indexed 138,283 scientific publications published between 2015 and 2020. ARCI's coverage is characterised by using the metadata available in scientific publications. First, I investigate the distributions of the indexed literature at various levels (research domains, countries, languages, open access). Articles make up nearly all the documents indexed with a share of 99% of ARCI. The Arts & Humanities and Social Sciences fields have the highest concentration of publications. Most indexed journals are published in Egypt, Algeria, Iraq, Jordan, and Saudi Arabia. About 8% of publications in ARCI are published in languages other than Arabic. Second, I use an unsupervised machine learning model, LDA (Latent Dirichlet Allocation), and the text mining algorithm of VOSviewer to uncover the main topics in ARCI. These methods provide a better understanding of ARCI's thematic structure. Next, I discuss how ARCI can complement global standards in the context of a more inclusive research evaluation. Finally, I suggest a few research opportunities after discussing the findings of this study.

## 1. Introduction

Arabic is one of the most widely spoken languages in the world and is used by more than 400 million people. Arabic was also the lingua franca during the Islamic Golden Age, serving as the language of science, poetry, literature, governance, and art. It played a catalytic role in developing scientific knowledge, building upon earlier traditions from Europe, China, Persia, India, and Africa. For centuries, science was done in several languages, until the rise and domination of English in the 20$^{th}$ century (Gordin, 2015).

Nowadays, scientific research is still a polyglot activity but research excellence is often equated to publishing in English in high-impact factors journals, as stated in the third principle of the *Leiden Manifesto* (Hicks et al., 2015). This is also stated in the *Helsinki initiative on multilingualism in scholarly communication*: “multilingual work should be fully acknowledged in scholarly assessments and English should not have more weight than other languages in communication” (The Committee for Public Information in Finland et al., 2019). Publishing in English has been largely discussed as a controversial topic in the literature (Aalbers, 2004; Coles, 1989; Garfield, 1989; Jiménez-Contreras & Ferreiro-Aláez, 1996; Pittler & Ernst, 2005; Short et al., 2001). This is problematic for the humanities and social sciences where research tends to be more engaged on national issues and published in local languages. Recent years have also seen a rapid development of new journals around the world. Identifying peer-reviewed journals of regional relevance and importance is a major issue for all scientific stakeholders (Moed et al., 2021). Such journals provide a valuable bibliometric data source on emerging issues. Protecting excellence in locally relevant scientific research is also key to preserve fields which have regional or national dimensions. Franssen and Wouters (2019) provide an extensive literature review of bibliometric publications that study the humanities between 1965 and 2018. In their review, Franssen and Wouters distinguish two periods in which the bibliometric system is configured in a variety of ways: 1965-1980s is characterized by bibliometrics embedded in a sociological theoretical framework. Whereas the 1980s-present period is marked by the embedding of bibliometric methodologies in the science policy and research evaluation contexts.

In 2020, the Arabic Citation Index (ARCI) was launched in the Web of Science (WOS) platform, first in Egypt and later in the rest of the Arab World. Clarivate Analytics partnered with the Egyptian Knowledge Bank (EKB), as part of the Egyptian Ministry of Education, to develop the first Arabic Citation Index. This launch is also part of Egypt's *Vision 2030* where knowledge, innovation and scientific research are key pillars to achieve scientific excellence (Egyptian Government, 2016). As mentioned by Dr Shawki, Minister of Education & Technical Education in Egypt and President of the EKB Project, the "aim is to work toward becoming a more knowledgeable Egyptian community that encourages learning as a part of everyday life. We look forward to building our economy and exporting our sciences globally in the Arabic language." (Clarivate Analytics, May 2018).

The focus of ARCI is on the scholarly research published in journals of publishers based in the 22 countries of the Arab League where Arabic is an official language (Algeria, Bahrain, Comoros, Djibouti, Egypt, Iraq, Jordan, Kuwait, Lebanon, Libya, Mauritania, Morocco, Oman, Palestine, Qatar, Saudi Arabia, Somalia, Sudan, Syria, Tunisia, United Arab Emirates and Yemen). Here, Arabic is "Modern Standard Arabic (MSA)", the formal written standard widely used in the media, education and scientific research throughout the Arab world. It is worth reminding that Arabic is a language with a diverse range of regional dialects. While these dialects are authentic native languages, they are rarely used in formal communication and are not standardised or taught in schools (Habash, 2010). MSA is based on the syntactic, morphological, and phonological features of Classical Arabic (Habash, 2010), which is the language of the Quran.

ARCI joins other regional citation databases in the WOS: the Chinese Science Citation Database, KCI - Korean Journal Database, SciELO Citation Index, and the Russian Science Citation Index. ARCI uses the same core features of WOS with a new Arabic language interface in addition to the classic English language interface. This new interface allows the user to run search queries in Arabic to find relevant records or researchers. The criteria for inclusion in ARCI are a subset of the Web of Science Core Collection criteria (Clarivate Analytics, 2019). The journals covered in ARCI are selected by a newly established Editorial board with members from Arab League countries who provide subject knowledge and regional insights. The selection process for ARCI is based on traditional scientific publishing standards and the scholarly research norms of the Arab region which include peer-review. However, the peer review methods such as single-anonymous and double-blind are not described. First, there is an initial triage to confirm content accessibility and format for all titles considered for indexation in ARCI. All journals must have an ISSN. Several elements are evaluated in this first step: journal title, publisher information, URL for online journals, content access, DOI/pagination and timeliness/volume. Next, the journals are reviewed from an editorial perspective. In this second step, each journal is evaluated to confirm it provides scholarly content, with a clear scope statement, article abstracts, cited references, content relevance with the stated scope or mission, quality of language consistent with scientific communications and an editorial board reflective of the field of the journal.

The indexing provided in ARCI aims to increase the exposure of Arabic-language research, allowing it to contribute to regional and worldwide research efforts. ARCI is a new addition to the WOS platform, and a separate subscription is required to access it. There is relatively little information available regarding this new database, which many scholars are still unfamiliar with. The content coverage of a database can be assessed from a variety of perspectives, including coverage of indexed publication sources, document types, disciplines and subject fields, publication language, and impact. An assessment of content coverage of a bibliometric database can only be made by conducting a large-scale analysis. Additionally, some features of a specific database and their potential uses can also influence its suitability for particular tasks such as research discovery or research evaluation. This study contributes to the literature in which metadata of publications is analysed bibliometrically to profile specific scholarly communities and publication practices (Franssen & Wouters, 2019). Therefore, the goal of this research is to describe the literature found in this new citation index. The main objective is to examine the regional research landscape from diverse perspectives. Such a study can help research managers and policymakers to better understand the regional research activity, by providing a more in-depth analysis of publication practices in a specific region.

The remainder of this paper is organized as follows. First, I provide a review of the scientometric literature on the inclusiveness of the Web of Science. Second, the data and methods used to conduct the analyses are expanded upon. Following that, I investigate a few content distributions at various levels (research domains, countries, languages, and open access). Next, I examine the primary subjects covered in ARCI by using the Latent Dirichlet Allocation model and the text mining algorithms of VOSviewer (van Eck & Waltman, 2010). Then, I discuss the role ARCI might play as a regional complement to global standards from a research evaluation perspective. Finally, I discuss the results of this study, identify its limitations, and suggest a few research directions.

## 2. Literature review

In the past 30 years, there have been some debates about the inclusiveness of bibliographic databases such as the Web of Science. For instance, Gibbs (1995) claimed that the Science Citation Index in WOS was biased toward Global North English-language scientific journals. Garfield (1997) responded that a statistically valid definition of bias was needed to conclude whether WOS was biased against so-called 'Third World' journals, referring to the law of concentration applied to science journals or Bradford law (Garfield, 1996). Later, Hicks (1999) discussed the difficulty to achieve full comprehensiveness of international social science literature and the bibliometric consequences this might have. One of her arguments was that the polyglot character of the social sciences might make them more difficult to cover in a single database.

During the past decade, many nations around the world such as Australia, the Czech Republic, Finland, Norway, Poland, Turkey, the UK, and many others have chosen to implement performance-based research funding (Aagaard, 2015; Hicks, 2012; Kulczycki, 2017; Tonta, 2017) and incentive schemes (Franzoni et al., 2011; Quan et al., 2017). Such incentives are mostly related to the publication activities of researchers (Rochmyaningsih, 2019), which are traditionally analysed by using multidisciplinary bibliographic data sources like the Web of Science, Scopus, Google Scholar, Dimensions and Crossref. These databases are all constructed in different ways, and hence all differ in terms of coverage of journals, document types, languages, disciplines and citation indexing. Such coverage differences have been the focus of various studies (Martín-Martín et al., 2021; Mongeon & Paul-Hus, 2016; Singh et al., 2021; Vera-Baceta et al., 2019; Visser et al., 2021). Mongeon and Paul-Hus (2016) find that English-language journals are overrepresented to the detriment of publications in other journals. They also show that the results of bibliometric analyses may differ depending on the bibliographic data source used. It has also been shown that research published in Social Sciences and Humanities mostly stays unnoticed when bibliometric sources like the WOS and Scopus are used for research (Aksnes & Sivertsen, 2019; Liu, 2017; Mongeon & Paul-Hus, 2016; Van Leeuwen et al., 2001). Moreover, as stated in the *Leiden Manifesto* (Hicks et al., 2015), research articles published in English are often considered to represent a high standard of quality. This has been studied in several European countries by Ochsner et al. (2018) and Sīle et al. (2018).

According to Sivertsen (2018), the usage of local language in scholarship is essential to promote interaction with stakeholders and the general public. This is also essential if science wants to fulfil its social obligations or have localised impacts (Garcia-Ramon, 2003; Hasse & Fischer, 2003; Huang, 2011; Samers, 2000). However, if evaluation regimes have an impact on publication practices and if they modify research agendas, researchers may decide to shift away from locally relevant research in favour to English-language audiences (Bianco et al., 2016). It is also worth reminding that different languages and communication venues affect different audiences (Hicks, 2004). Non-English journals serve communication functions that are distinct from those of mainstream English journals, as demonstrated by Chavarro et al. (2017): they provide researchers with opportunities for initiation into scientific publication and they address topics that might be underrepresented in mainstream publication titles.

Such issues have been discussed by several groups who have set multiple initiatives to enhance research evaluation. The *San Francisco Declaration on Research Assessment* (https://sfdora.org) emphasises that the publications' scientific content are more significant than the journals' publication metrics. Wilsdon et al. (2015) argue that evaluation should promote the diversity and plurality of research in *The Metric Tide* report. In the *Leiden Manifesto*, authors call to protect excellence in locally relevant research (Hicks et al., 2015). More recently, the Helsinki Initiative on Multilingualism in Scholarly Communication has been launched to encourage the dissemination of research findings in all languages (The Committee for Public Information in Finland et al., 2019). In that sense, regional or national databases, that are created to comprehensively cover all subjects and languages, are therefore crucial to develop a more balanced multilingualism in scholarly communication. The need to create national citation indexes was also discussed by Pislyakov (2007). There are several examples of such databases created to cover non-English scientific literature in Brazil (Packer et al., 1998), China (Jin & Wang, 1999; Su et al., 2014; Ye, 2014), India (Yadav & Yadav, 2014), Japan (Negishi et al., 2004), Korea (Seol & Park, 2008), Russia (Moskaleva et al., 2018), Serbia (Pajic, 2015), or in Taiwan (Chen, 2004). The common objective of these developments is to provide more visibility and an easier access to journals publishing scientific papers in languages other than English.

There is also research on how well various bibliographic databases are able to provide a global complete coverage of the scientific literature, with special attention paid to the proper coverage of journals published in languages other than English or in countries of the Global South. Garfield (1995) reminded that the Science Citation Index (SCI) and other ISI's products were selective, hence they were not comprehensive in terms of coverage of scientific journals published globally. Chavarro et al. (2018) explored the extent to which the indexation in Web of Science might be an indicator or quality. They found that journals with comparable features and editorial requirements were often treated differently because of their publication country, field of study, and language. They warned research evaluators and joined other authors in urging caution in terms of research evaluation (Alperin, 2014; Garfield, 1995; Mounier, 2018): indeed, a multidimensional picture of local research would be obtained by including regional or local journals.

In a recent study, Brasil (2021) gave an overview of Brazilian papers indexed in regional databases. He found that publications not included in WOS are primarily written in Portuguese, with a considerable share indexed by regional databases and covering subjects that are not addressed in WOS. Brazilian scholarship includes not only papers published in prominent international journals, but also regionally pertinent topics that are mostly written for a Portuguese-speaking readership as well. He demonstrated that, although integrating international metrics from well-established databases could seem like a decent and straightforward way to improve the local science system, databases like WOS do not provide the whole story. He also concludes by arguing that research evaluators should aim at developing a more comprehensive assessment framework to capture the complexity of local science by including regionally relevant databases.

## 3. Data and Methods

### *3.1. Data*

ARCI has a coverage back to 2015. ARCI data was extracted on May 11th, 2022. Both 2021 and 2022 records were excluded since these years were not completely indexed yet in the database, and 2021 publications are expected to be fully indexed by the end of 2022. Full records and cited references were exported from the WOS platform. The dataset under study consists of 138,283 records for the 2015-2020 period. ARCI more than doubled in size in terms of number of indexed records since October 2020, when it was indexing 65,208 records for the 2015-2019 period (El-Ouahi, 2021). As a result of this expansion, it is now possible to examine a wider corpus of Arabic-language scientific literature. Figure 1 presents the number of records by publication year indexed in ARCI.

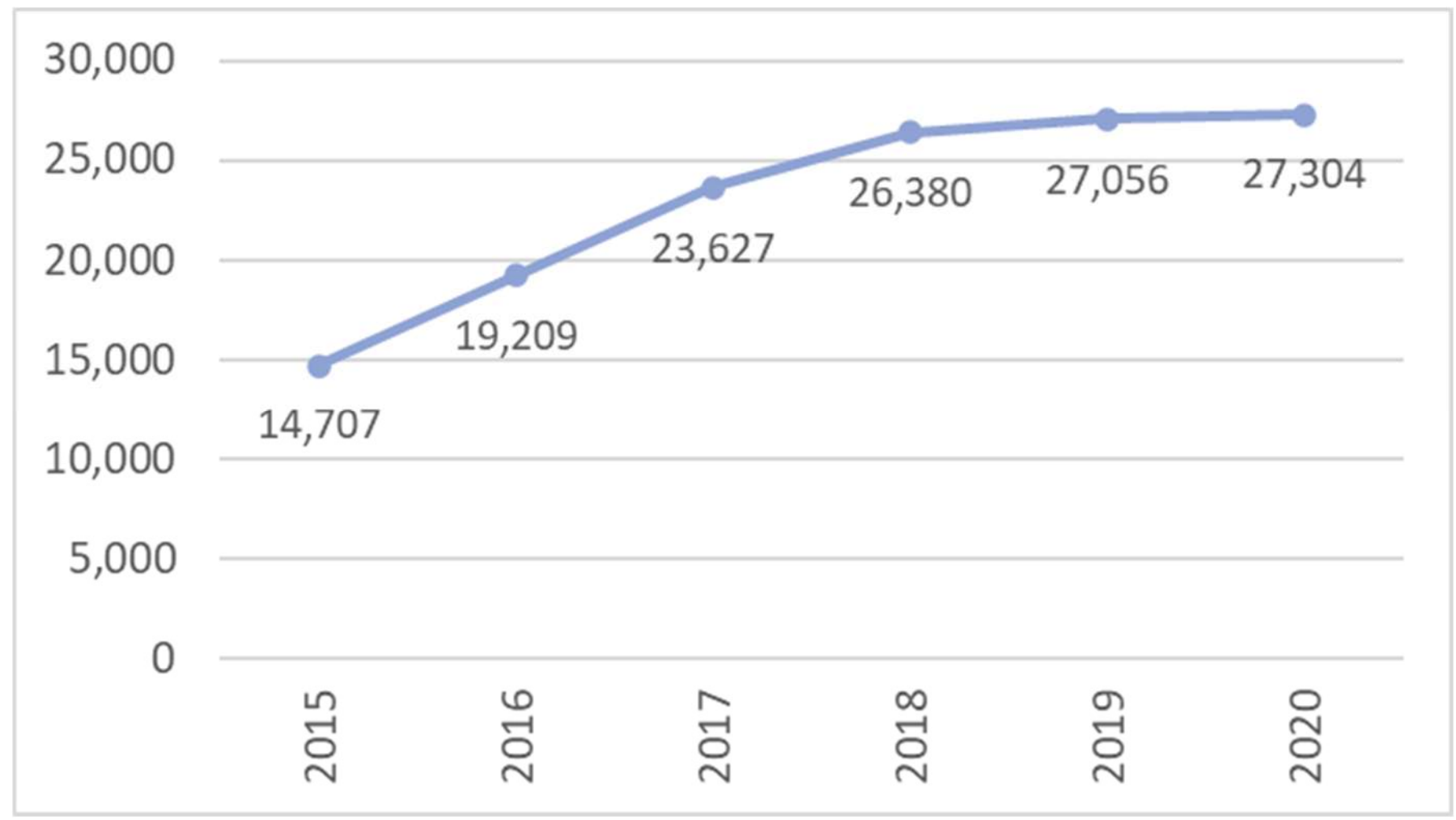

**Fig. 1. Number of ARCI records by publication year (2015-2020)**

This database is well structured with 48 fields of information in each record allowing multiple bibliometric analyses (e.g. Publisher Information, Funding Information, Research Area, Open Access Indicator, Cited References, Citations, Usage Counts, ESI Highly Cited Paper/Hot

Paper). In addition to essential metadata available in English as in the Web of Science Core Collection, ARCI has some specific information written in Arabic such as *authors names*, *article title*, *publication name*, *author keywords*, *abstract* and *author address.* ARCI records also show the ARCI times cited and the Total Times Cited Count (Web of Science Core Collection, Arabic Citation Index, BIOSIS Citation Index, Chinese Science Citation Database, Data Citation Index, Russian Science Citation Index, SciELO Citation Index) as well as the Cited References and the Cited Reference Count.

### *3.2. Methods and limitations*

In this study, I use bibliometric methods to characterize the literature indexed in ARCI. The objective is to examine the Arabic research landscape from various perspectives. Such analyses can help research managers and policy makers to better understand research activity in this part of the world. They can also provide a more detailed understanding of publication practices in the region. I conducted a bibliometric analysis to study the research output indexed in ARCI as follows.

A series of features can be used to profile journals. For instance, the country of the journal publisher, its editor, reviewers, authors and readers can be analysed to understand the geographical focus of ARCI. First, a journal distribution across countries was determined. In this study, the publisher's country information is used to determine the geographical distribution of journals indexed in ARCI. However, it is important to be aware that a considerable number of journals would be edited and published in different countries.

Next, other aspects of the indexed literature were explored such as the distribution of publications by research fields, languages, and access types. Then, I analysed the scientific research output at the country level by using the authors addresses. A full counting method is applied in this study to report the number of publications by authors' countries. I used the following approach to assign a country to each address: authors' addresses have a common structure which consist of several elements including the institution name, the college name, the department or laboratory name, the city, and the country. 74,283 addresses were found in ARCI. In some addresses, neither the country nor the city is available. For such cases, when available, I assigned the country of the addresses sharing the same institution name. As a result of this approach, a country was assigned to 95% of all the addresses found.

Finally, other aspects analysed in this paper relate to the topics addressed in ARCI. I used topic modelling algorithms and text mining techniques to describe the topics discussed in the publications making up ARCI. Such methods rely on statistical analysis of the words in such publications, identifying clusters of co-occurring words, and detecting the topics discussed and the relationships between them. Many machine learning algorithms have been developed to understand, group or search information from large text databases. In natural language processing, a topic model is a statistical model to discover the hidden semantic structures or topics that occur in a collection of documents. There are several models available, such as Latent Semantic Analysis (LSA) (Landauer et al., 1998), Probabilistic Latent Semantic Analysis (PLSA) (Hofmann, 1999), Latent Dirichlet Allocation (LDA) (Blei et al., 2003); as well as some derived models from the latter, like the Pachinko Allocation (Newman & Block, 2006) or the Relational Topic Modelling (Chang & Blei, 2009).

LDA model is probably the most well-known and commonly used model. It has been proposed by Blei et al. (2003) to classify documents into topics. LDA is a generative probabilistic model of a corpus. The basic idea is that publications are composed of groups of words with no sequential relationship between them. As documents can include multiple topics, each record can be described by a distribution of topics. And each topic is characterized by a distribution over words, described as a distribution of terms in a fixed vocabulary. LDA can be used to identify a group of topics, assign a group of words to a topic and determine the mix of topics in each publication.

The LDA model has been frequently used to examine the structure of an aggregated literature, in different fields, such as in the automated analysis of abstracts of academic articles (Griffiths & Steyvers, 2004) and in the analysis of blogs content (McCallum et al., 2007; Nallapati & Cohen, 2008). It has also been used in the study of content on Twitter (Weng et al., 2010) and to recommend academic publications (Jiang et al., 2012). More recently, it has been applied for the detection of topics in large collections of press articles (Lee et al., 2015).

Previous studies have shown that LDA performed well to understand the topical structure of a scientific corpus (Han, 2020; Suominen & Toivanen, 2016; Yau et al., 2014). Although LDA can produce excellent estimation results, it has two main specificities. First, determining the topic correlations between each of the topics is difficult. Second, before applying LDA, one must define the number of topics to model the corpus which is typically unknown in advance. Determining the natural number of topics is a controversial issue (Arun et al., 2010). Although various computational approaches have been proposed to optimize the number of topics (Griffiths & Steyvers, 2004; Zhao et al., 2015), several authors argue that human judgment is the best way to define the number of topics (Graham & Milligan, 2012; Newman & Block, 2006). The latter approach is the one used in this study.

Since VOSviewer does not support the Arabic language, LDA is applied to ARCI as follows. This research was limited to words written in the Arabic alphabet. Because there was no access to the full texts of the publications and since the titles, abstracts and keywords summarise the full contents of publications, the analysed corpus of documents consists of combinations of words available in the title, abstract and author keywords of each of the 138,283 records downloaded from ARCI. Normalization of characters (uppercase/lowercase letters) was applied to the corpus. Stemming and lemmatization procedures were applied and stop words were removed based on a standard list of stop words for Arabic (Brahmi et al., 2012). There are many publicly available tools for LDA. Here, I applied the *LdaModel* available in *Gensim*, a well-known Python programming library widely used in unsupervised machine learning.

With regards to the VOSviewer term mapping, the analysis is limited to words written in the Roman script languages. At the very least, the title of each record in ARCI is also written in English and 84% of the records have an abstract written in English. VOS viewer has basically two limitations. The first one is imposed by the data which includes some noise. Authors make some choice when selecting the words and terminology used in their publications which might include the usage of synonyms/homonyms not recognised by VOSviewer. The second limitation is related to the loss of information when terms are projected on a two-dimension Euclidean space. Also, the map includes specialised terms and more general ones which can be used in various contexts. Asa result of these limitations, a map interpretation should always be done with caution.

## 4. Results

First, I analyse the research domains in ARCI by number of records and the proportion they represent in the database. Next, I present the journals distribution by countries. Then, the most productive countries are examined, followed by an analysis of the languages of publications and their access types. Finally, I focus on the main topics covered in the Arabic scientific literature indexed in ARCI.

*4.1. Research areas distribution*

*Research areas* constitute a subject categorisation scheme that is shared by all WOS product databases. This is particularly helpful when analysing documents from multiple databases related to the same research areas. All 153 research areas in the WOS are grouped into five broad categories: Arts & Humanities, Life Sciences & Biomedicine, Physical Sciences, Social Sciences, and Technology.

I relied on the journal category and not on the topics covered in the individual publications to analyse the disciplinary coverage in ARCI. These categories or areas, which are defined at the journal level, are used as proxies for scientific fields. The ARCI records relate to 21 research areas in the dataset under study. Currently, 23,864 records (around 17% of ARCI), do not contain data in the Research Area field. In Figure 2, I summarise the share of records by research area in the database. I have limited the analysis to the 15 research areas with a share higher than 1%. Film, Radio & Television, Communication, Sociology, Social Work, Geography, Music, Psychiatry and Demography all have a share below 1%.

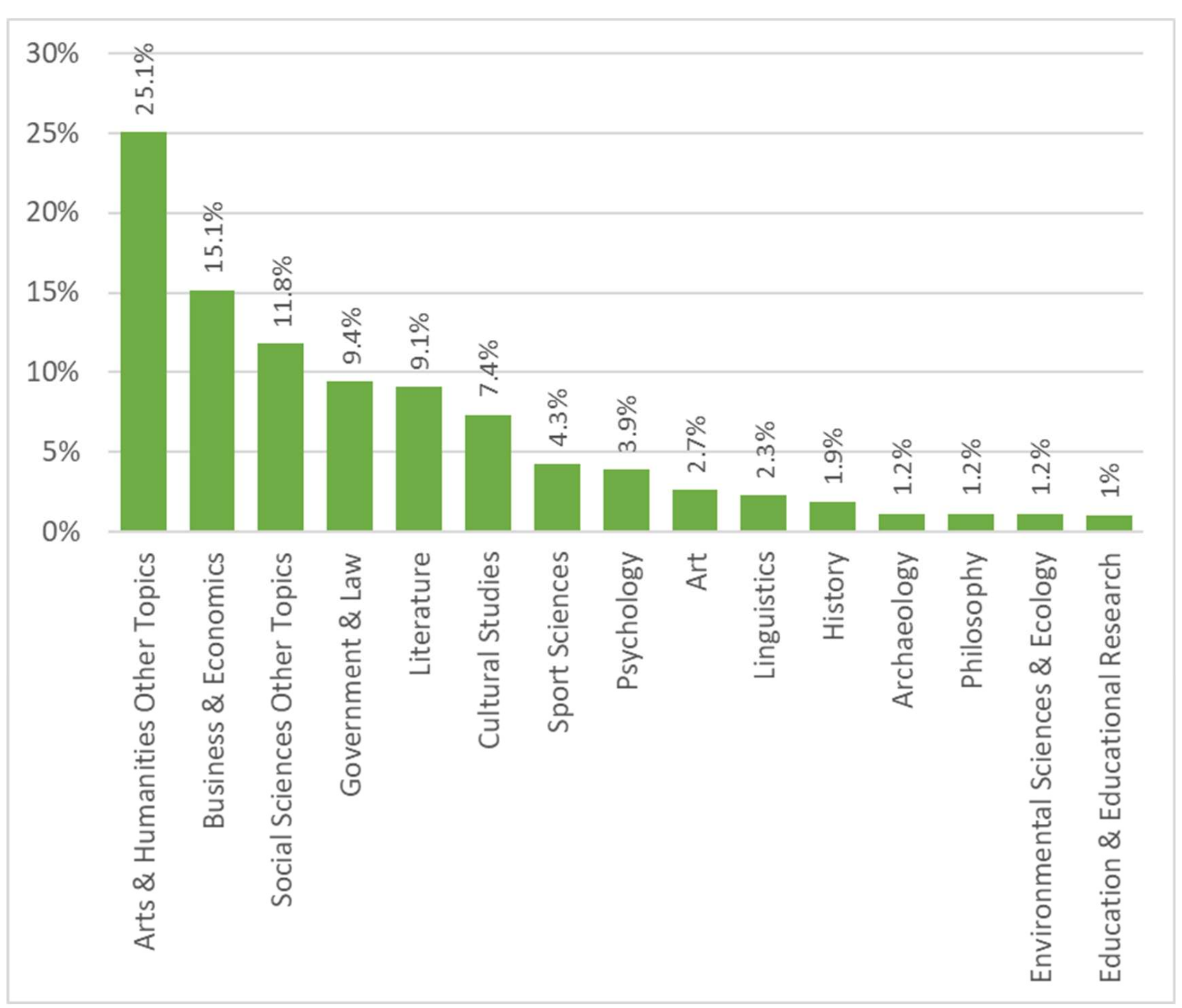

**Fig. 2 Share (%) of documents by research areas in ARCI (2015-2020)**

I have also summarised the shares of the number of papers within each of the five broad domains in Figure 3.

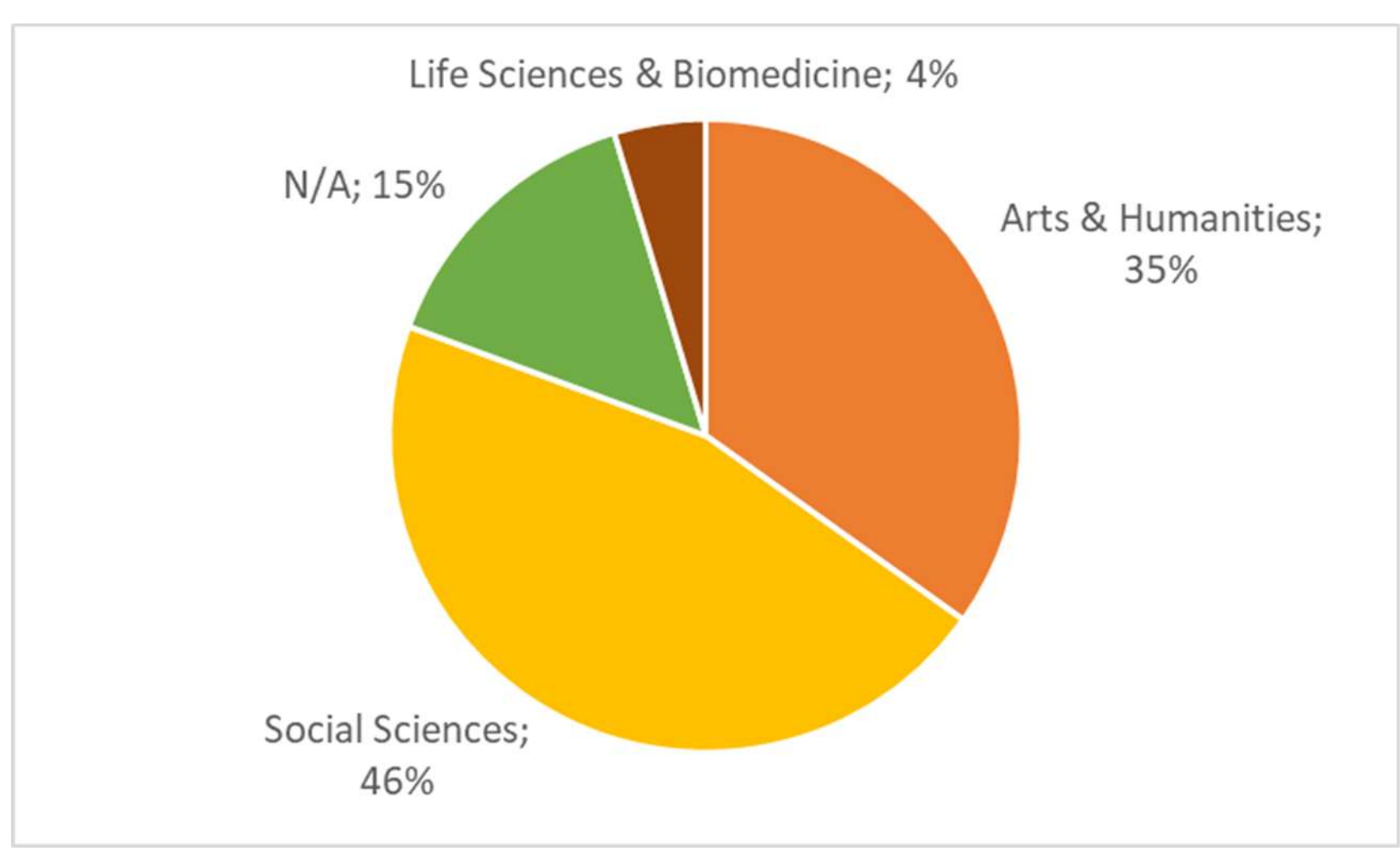

**Fig. 3 Share (%) of documents by broad categories in ARCI (2015-2020)**

This figure shows ARCI contains mainly journals in the Arts & Humanities and Social Sciences categories. These categories represent 81% of ARCI total coverage. Journals in Life Sciences & Biomedicine account for 5% of the coverage. As mentioned earlier, 15% of records retrieved do not contain information about the research area. It is worth noting there are no journals related to Technology or Physical Science categories. This confirms the current focus of ARCI. Regional issues in Arts & Humanities as well as Social Sciences dominate the ARCI coverage.

ARCI also offers its own research categories. When analysing the records with the ARCI classification, only 243 records do not contain information about the research categories, representing less than 0.18% of the total database. The alluvial diagram in Figure 4 shows the numbers of records as per 35 ARCI's own research categories, represented by the sizes of the nodes, along with the corresponding numbers as per 23 WOS research areas.

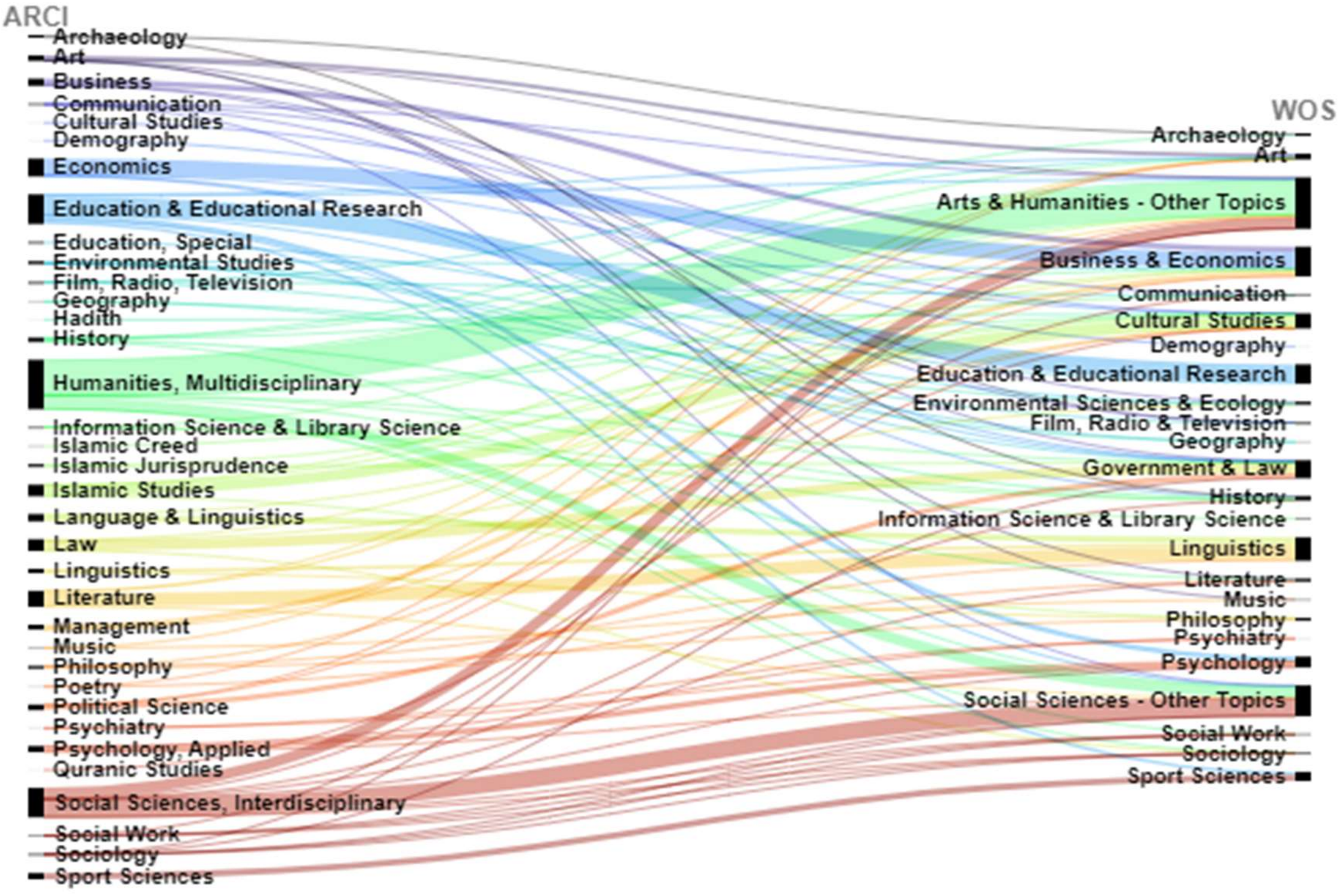

**Fig. 4 Number of documents by research areas in ARCI (2015-2020) as per ARCI's categories (left) and corresponding WOS research areas (right)**

We retrieve similar results, however, some differences emerge. The *Business & Economics* category in WOS is broken down into three categories in ARCI: *Business, Economics* and *Management*. The *Cultural Studies* in WOS consists of *Language & Linguistics* and the following categories in ARCI: *Islamic Studies, Islamic Jurisprudence*, *Islamic Creed*, *Quranic Studies* and *Hadith* which are fields well studied in the Arab region. There is also a distinction between *Literature* and *Poetry* in ARCI, which are both classified as *Literature* in WOS. Finally, *Law & Political Science* in ARCI are both categorized as *Government & Law* in WOS.

*4.2. Content coverage by publisher's country*

In this section I analyse the coverage by country. First, I examine the types of documents indexed in ARCI. Table 1 lists the number of documents per type and they share they represent in the database.

**Table 1. Number and share of ARCI records by document type (2015-2020)**

| *Document Type* | *Records* | *Share (%)* |
|---|---|---|
| Articles | 136,819 | 98.94 |
| Review Articles | 911 | 0.66 |
| Editorial | 174 | 0.13 |
| Other | 171 | 0.12 |
| Art and Literature | 117 | 0.09 |
| Bibliographies | 88 | 0.06 |
| Meeting | 3 | - |

ARCI is primarily composed of journal articles. Close to 99% of documents indexed are articles. Other document types all represent less than 1% of the database.

Now, I focus on the distribution of journals over countries published in the Arab League countries. As mentioned earlier, each journal is assigned to a country based on the country in which the publisher is located. But, before analysing the country distribution in ARCI, I examined the coverage of Arab journals in the various citation indices in the Web of Science Core Collection (WOS CC): Science Citation Index Expanded (SCIE), Social Sciences Citation Index (SSCI), Arts and Humanities Citation Index (AHCI) and Emerging Sources Citation Index (ESCI). This coverage is presented in Figure 5. ESCI was launched in 2015, with backfiles dating back to 2005. It covers all disciplines with international and broad scope publications as well as regional or specialty area focus. ESCI differs from SCIE, SSCI and AHCI in terms of process used by Clarivate to select journals. This process uses a set of 28 criteria to evaluate journals. These criteria are split into 24 quality criteria (editorial rigor and best practice at the journal level) and four impact criteria to select the most impactful journals in their field by using citation activity as the main indicator of impact. Journals that meet the 24 quality criteria are indexed in ESCI, and journals that meet the four additional impact criteria enter SCIE, SSCI or AHCI based on their subject category [1].

As of October 2020, 21,419 journals were indexed in WOS CC. 144 journals (or 0.67%) are published in 13 of the 22 Arab League countries: 66 in SCIE, 1 in SCIE and SSCI and 77 in ESCI. Out of these 144 journals, 134 (93%) are published in English only. The remaining 10 journals (7%) have published papers in several languages during the study period: English (78%), French (12.6%), Arabic (5.8%), Spanish (3.4%), Afrikaans (0.05%) and Italian (0.05%).

[1] https://clarivate.com/products/scientific-and-academic-research/research-discovery-and-workflow-solutions/web-of-science/core-collection/editorial-selection-process/editorial-selection-process/

The United Arab Emirates (UAE), Egypt and Saudi Arabia are the three most represented Arab countries in WOS CC with a total of 113 journals and a share of 78% of all journals published in the Arab region and indexed in WOS CC.

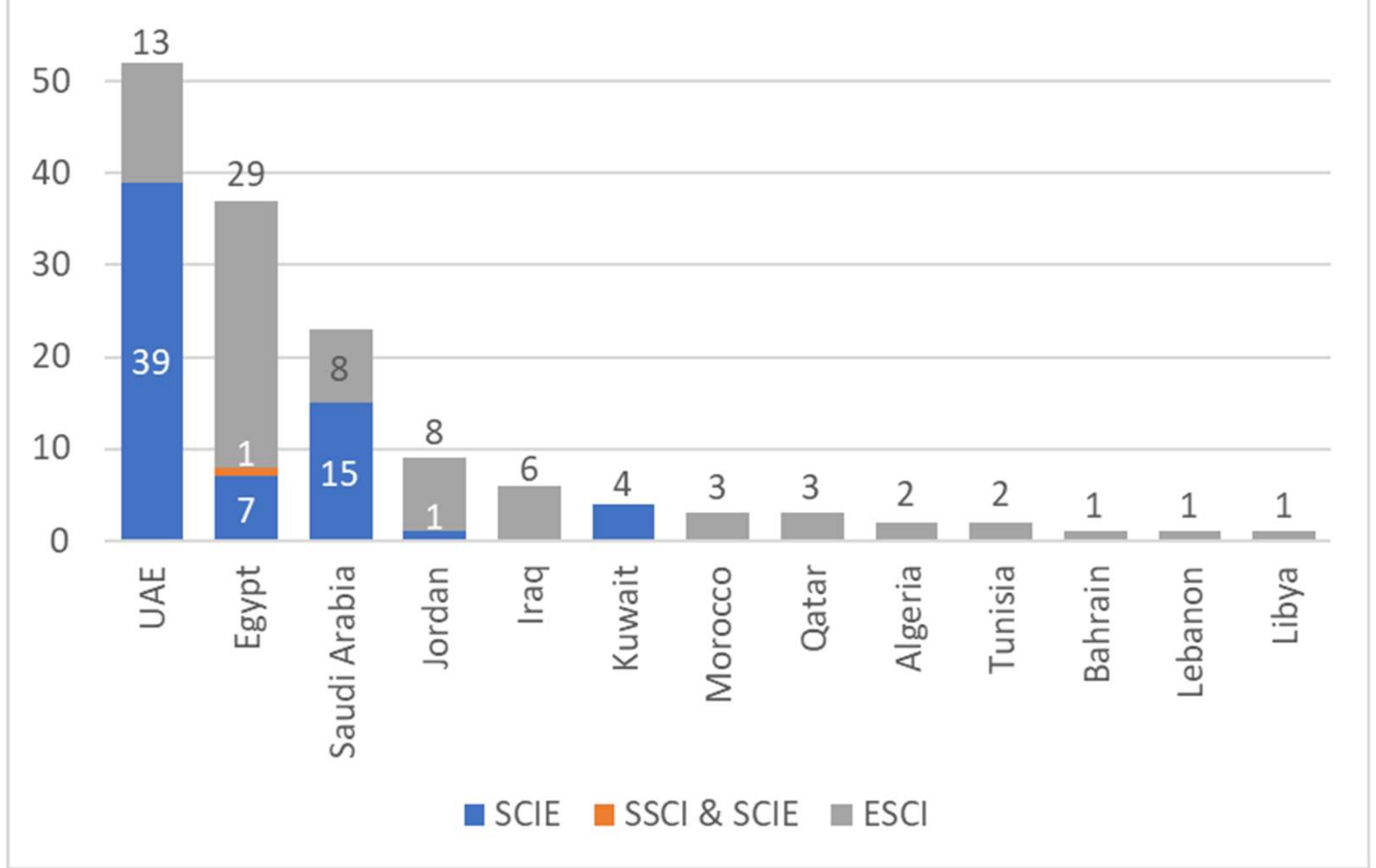

**Fig. 5 Number of journals by Arab country in WOS CC citation indices (September 2020)**

Although the criteria for inclusion in ARCI are a subset of the Web of Science Core Collection selection process, there is no overlap between WOS CC and ARCI. The distribution by country of publisher in ARCI is represented in Figure 6. ARCI indexes content from 19 of the 22 Arab League countries. Content for Djibouti, Comoros and Somalia is not indexed yet. As of June 2021, 613 journals were indexed in ARCI. Journals published in Egypt, Algeria, Iraq, Jordan and Saudi Arabia represent 83% of the journals indexed in ARCI. Again, this analysis does not take into account the location of the editor but the publisher's country. There might also exist journals publishing scientific literature in Arabic but located in countries not part of the Arab League.

Egypt and Algeria cumulate more than 60% of the journals indexed in ARCI. There is currently a high concentration of journals published from these two countries in ARCI. Then, 22.3% of the ARCI journals are published in Iraq, Jordan and Saudi Arabia with respective shares of 11.1%, 6.0% and 5.2%. Such concentration might be due to several reasons such as publishers' awareness and readiness in specific countries to provide their journals' data for indexation.

The submission process is managed through the Egyptian Knowledge Bank website (www.arcival.ekb.eg) and journals are evaluated by an independent editorial board according to the ARCI selection process as explained in the introduction of this paper. Publishers in countries which have set up national journal platforms and initiatives might also have an advantage in providing journals' data more easily as per publishing standards and selection criteria of indexing databases. For example, the Algerian Scientific Journal Platform (ASJP) (www.asjp.cerist.dz) has been developed by the Ministry of Higher Education & Research in Algeria. ASJP consists of 757 journals and more than 187,000 articles. Another similar initiative is the Iraqi Academic Scientific Journal platform (www.iasj.net) set up by the Ministry of Higher Education & Scientific Research of Iraq. IASJ currently lists more than 218,000 articles published in 361 open access peer-reviewed journals by 92 Iraqi universities and research institutions. Similarly, the Ministry of Higher Education and Research in Morocco developed the portal of Moroccan scientific journals (https://revues.imist.ma) which includes 186 scientific journals. The common goal of all these initiatives is to improve the visibility of local journals by improving their publishing standards.

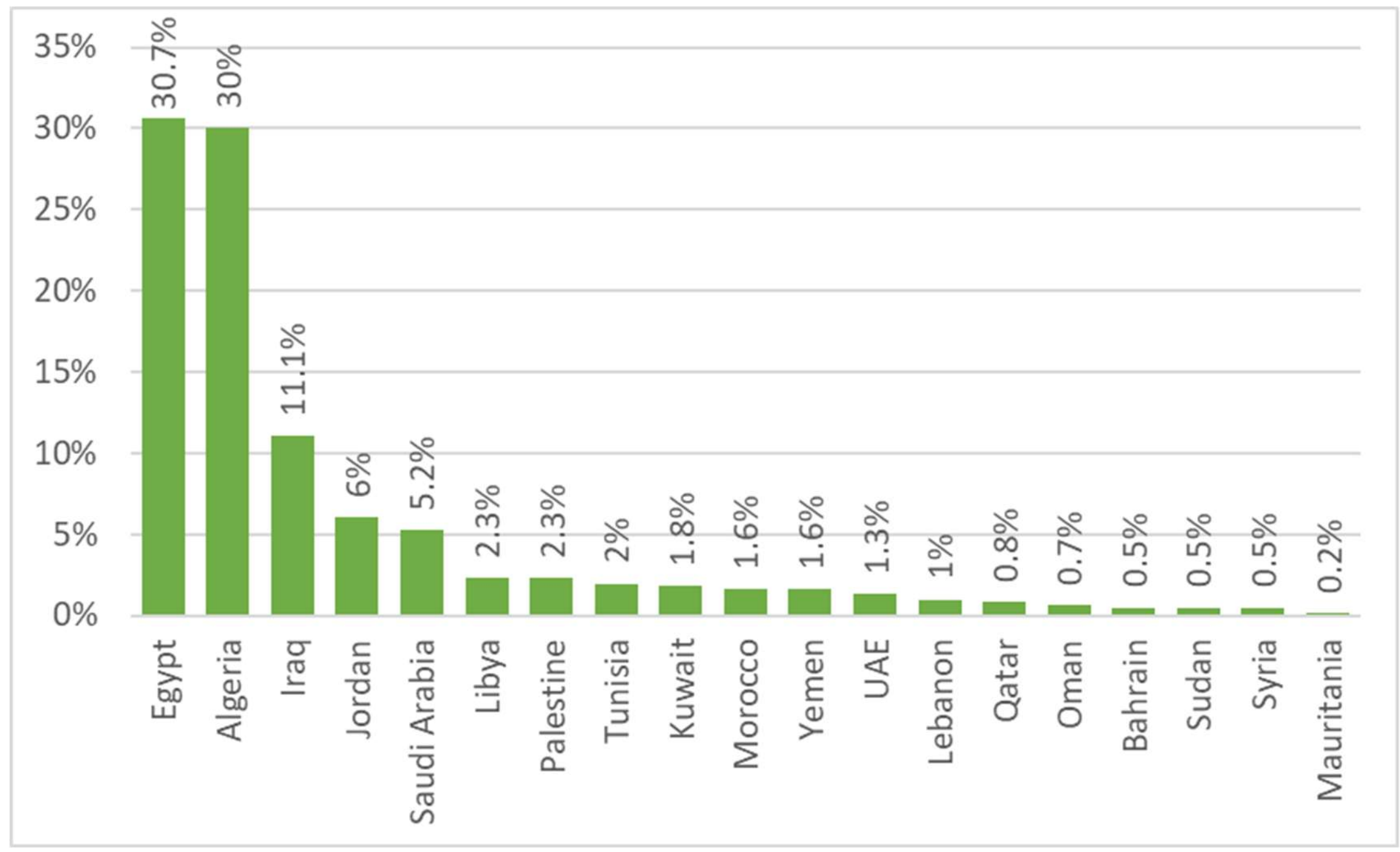

**Fig. 6 Share (%) of journals by country in ARCI (2015-2020)**

The heatmap in Figure 7 shows the contribution to each research category in terms of number of journals by country of publisher in ARCI between 2015 and 2020. Such visualisation is useful to understand the level of contribution and specialisation of each country. Egypt, Algeria, Iraq and Saudi Arabia contribute to most research categories. While Egypt has the highest contribution in Education (41 journals in ARCI), Algerian publishers contribute the most to scientific research in Business and Economics (47) and publishers in Iraq show a higher level of focus on Humanities (23). We notice publishers in Algeria also contribute the most to Social Sciences (30), Humanities (26), Language & literature (26 and Law & Political Science (25).

| | Business & Economics | Humanities | Social Sciences | Education | Language And Literature | Law And Political Science | Philosophy And Religion | Multidisciplinary Sciences | Sport Sciences | Islamic Studies | Art And Architecture | Communication, Journalism, Information Science | Science, Technology, Engineering & Medicine | Archaeology | Agriculture | Architecture | Geography | Hospitality, Leisure, Sport & Tourism | **Total** |
|---|---|---|---|---|---|---|---|---|---|---|---|---|---|---|---|---|---|---|---|
| Egypt | 20 | 12 | 21 | 41 | 18 | 4 | 20 | 11 | 9 | 6 | 8 | 8 | 4 | 4 | 1 | | | 1 | 188 |
| Algeria | 47 | 26 | 30 | 4 | 26 | 25 | 9 | 5 | 3 | 3 | | 2 | | 3 | | 1 | | | 184 |
| Iraq | 9 | 23 | 3 | 3 | 6 | 6 | 2 | 3 | 5 | | 1 | 1 | 4 | | 1 | | 1 | | 68 |
| Jordan | 8 | 6 | 4 | 4 | 3 | 1 | 4 | 5 | | 1 | 1 | | | | | | | | 37 |
| Saudi Arabia | 5 | 6 | 3 | 4 | 2 | | 1 | 2 | | 1 | 2 | 1 | 3 | 2 | | | | | 32 |
| Libya | 2 | 5 | 1 | | 1 | | | 3 | 1 | | 1 | | | | | | | | 14 |
| Palestine | 2 | 7 | | 3 | | | | 2 | | | | | | | | | | | 14 |
| Tunisia | | 2 | 3 | | 1 | 1 | 2 | | | 1 | | 2 | | | | | | | 12 |
| Kuwait | 2 | 2 | 2 | 3 | | | | | | 1 | 1 | | | | | | | | 11 |
| Morocco | | 1 | 1 | | 3 | 4 | 1 | | | | | | | | | | | | 10 |
| Yemen | 1 | 2 | 1 | 3 | 1 | | 1 | | | 1 | | | | | | | | | 10 |
| UAE | | | 1 | 2 | | 2 | | 1 | 1 | 1 | | | | | | | | | 8 |
| Lebanon | | 2 | | | | 1 | 1 | 1 | | | | | 1 | | | | | | 6 |
| Qatar | | | | | | 1 | 1 | | | 3 | | | | | | | | | 5 |
| Oman | 1 | | | 1 | 1 | | | | | | 1 | | | | | | | | 4 |
| Bahrain | 1 | 1 | | 1 | | | | | | | | | | | | | | | 3 |
| Sudan | | 1 | 1 | | 1 | | | | | | | | | | | | | | 3 |
| Syria | | | | 1 | 1 | | | | | | | 1 | | | | | | | 3 |
| Mauritania | | 1 | | | | | | | | | | | | | | | | | 1 |
| **Total** | 98 | 97 | 71 | 70 | 64 | 45 | 42 | 33 | 19 | 18 | 15 | 15 | 12 | 9 | 2 | 1 | 1 | 1 | 613 |

**Fig. 7 Number of journals by publisher's country and research category in ARCI (2015-2020)**

ARCI is still new and is still growing. Considering, the initiatives taken by governments to improve the visibility and the publishing standards of local journals, it will be interesting to analyse how this new citation index will evolve over time in terms of coverage by journals' countries and research category.

*4.3. Languages coverage*

Table 2 shows the coverage of records in terms of language of publications in ARCI. Arabic obviously dominates the database with 126,968 publications, representing a share of around 91.8%. As ARCI aims to provide more exposure to journals published in the Arab League countries, it is no surprise to see Arabic as the dominant language in this database. The second most represented language is English with 7,849 records (5.68%) then followed by 2,960 publications in French (2.14%). 506 publications in 10 other languages represent 0.37% of this database.

56 records have an "unspecified" language in the Web of Science, which consist of 55 publications in Kurdish and one in Amazigh. The languages of these records were found by identifying the language used in the full text of the related publications. Today, the two principal written Kurdish dialects are Kurmanji and Sorani. Along with Arabic, Sorani is one of the two official languages of Iraq and is simply referred to as "Kurdish" in political documents. The Amazigh language, also known as Tamazight, is widely spoken in Northern Africa. It became an official language of Morocco in 2011 along with Arabic. Arabic and Tamazight are also the official languages of Algeria since 1963 for the former and since 2016 for the latter.

**Table 2. Number and share of records by language in ARCI (2015-2020)**

| *Language* | *Records* | *Share (%)* |
|---|---|---|
| Arabic | 126,968 | 91.82 |
| English | 7,849 | 5.68 |
| French | 2,960 | 2.14 |
| Spanish | 124 | 0.09 |
| German | 108 | 0.08 |
| Persian | 61 | 0.04 |
| Kurdish | 55 | 0.04 |
| Hebrew | 47 | 0.03 |
| Italian | 46 | 0.03 |
| Russian | 38 | 0.03 |
| Turkish | 15 | 0.01 |
| Chinese | 11 | 0.01 |
| Amazigh | 1 | - |

It is also worth reminding many journals indexed in ARCI provide publication in multiple languages. Several countries from the Arab League are former British or French colonies which explains why English and French are the main non-Arabic languages in ARCI. The presence of English is also not surprising since many local journals use English as their language of publication in order to reach a larger community. Other languages may suggest research published in ARCI journals might also tackle regional issues of interest with neighbour countries.

### *4.4. Co-authorship structure and research output by country*

Analysing the research output at the authors' address level is also particularly useful to understand the regional publication practices and how they relate to the social structures that we find in specific research fields such as the Humanities.

The first element analysed here is the co-authorship structure found in ARCI. In research evaluation and management, co-authorship information is often used to characterise scientific collaboration relations (Glanzel, 2001). Table 4 reports the distribution of records in ARCI by number of authors along with the share of the database they represent between 2015 and 2020.

As presented in Table 3, the most common type of authorship in ARCI is single with a share of about 66%. This shows that there is a marked preference for single work. This is not surprising considering single authorship is a common practice in humanities and social sciences which represent a high share of ARCI. Next, double and triple authorship publications represent respectively about 25% and 6.8% of ARCI. Less than 1.6% publications indexed in ARCI are co-authored by 4 or more authors.

**Table 3. Distribution and share of records by number of authorships in ARCI (2015-2020)**

| *Number of authors* | *Records* | *Share (%)* |
|---|---|---|
| 1 | 91,574 | 66.22 |
| 2 | 35,029 | 25.33 |
| 3 | 9,461 | 6.84 |
| 4 | 1,846 | 1.33 |
| 5 | 330 | 0.24 |
| 6 | 30 | 0.02 |
| 7 | 9 | 0.01 |
| 8 | 1 | - |
| 9 | 2 | - |
| 11 | 1 | - |

In Figure 8, the focus is on the authorship structure by research category.

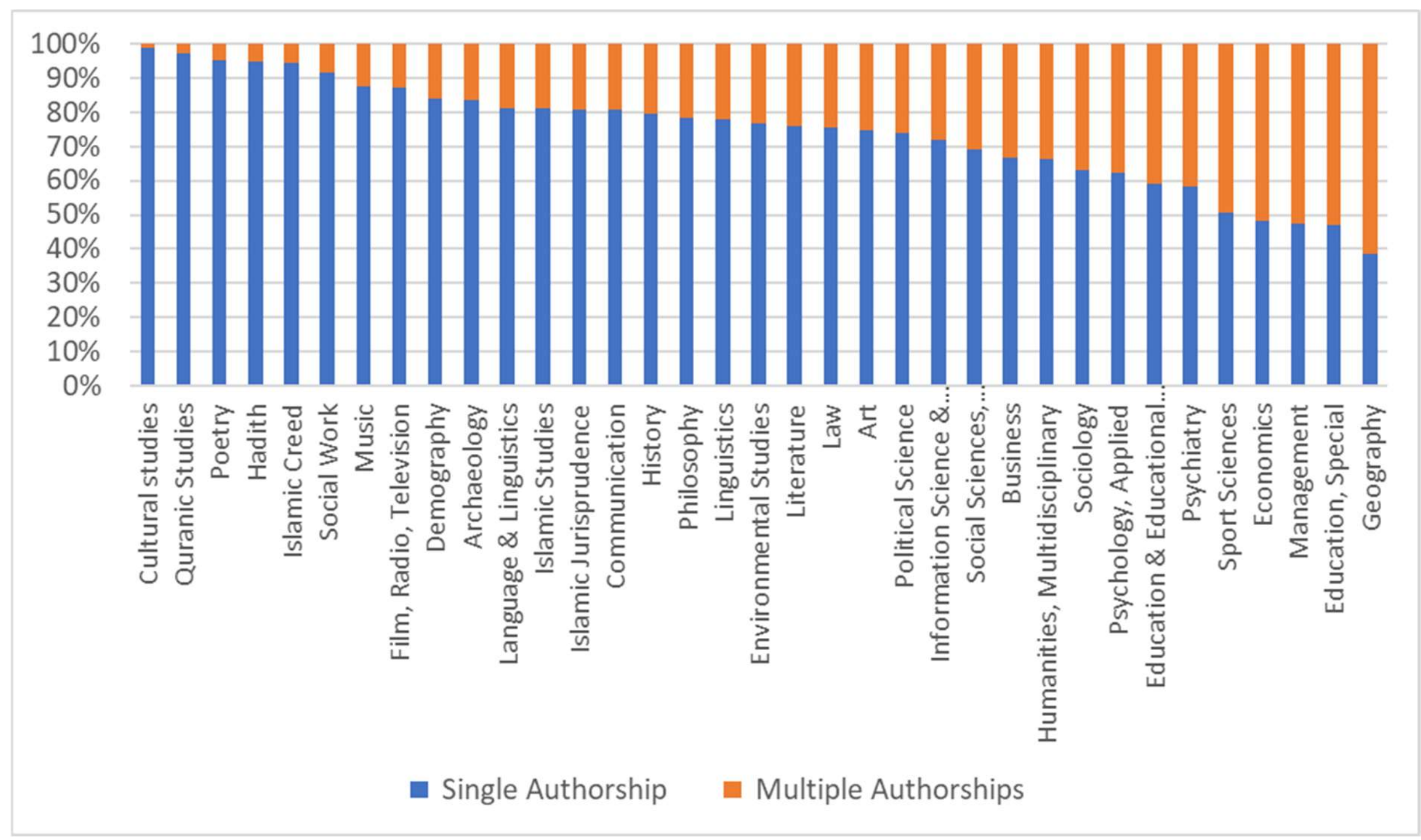

**Fig. 8 Share of single and multiple authorships by research category in ARCI (2015-2020)**

In some research fields, single authorship is rather the norm while in others collaborative work is more frequent. In the first case, Cultural Studies, Quranic Studies, Poetry, Hadith, Islamic Creed and Social Work all show a share of single authorship publications higher than 90%. On the other hand, multi authorship is more frequent in Geography (61%), Special Education (53%), Management (52%), Economics (51%), Sport Sciences (49%), Psychiatry (41%) and Education & Educational Research (40%). These results suggest that those areas exhibit a more collaborative aspect.

The second element analysed in this section is the research output by country based on the authors' affiliation(s). Figure 9 presents the number of records indexed in ARCI by authors' country for countries with more than 100 records assigned to them. Algeria dominates with close to 33,000 publications. Egypt, Iraq, Saudi Arabia and Jordan then follow. Most countries in Figure 9 are part of the Arab League but some cases stand out such as Iran, Malaysia, United States of America, France, Turkey and United Kingdom.

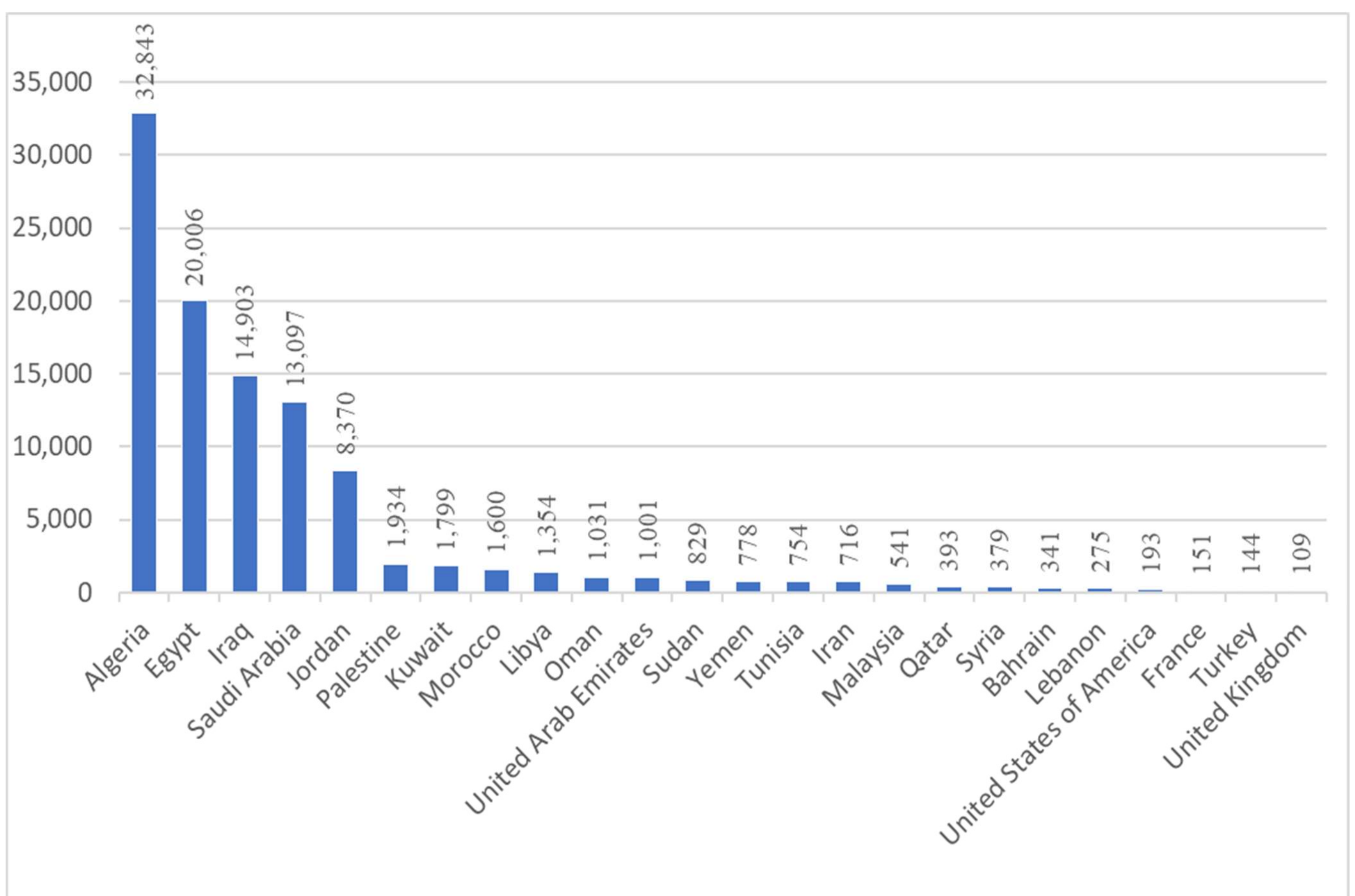

**Fig. 9 Number of records indexed in ARCI by country for countries with more than 100 publications (2015-2020)**

*4.5. Main topics*

As mentioned earlier, before applying LDA, one must define the number of topics for the corpus. One option is to examine the performance of text clustering on a small dataset. Another way is to choose the number of topics based on judgments or tests (Blei et al., 2003). Similarly to the term map obtained in section 3.6, the corpus is organized into 7 topics which are listed in Table 4.

As mentioned earlier, the model is applied on the combinations of words available in the title, abstract and author keywords of all records indexed in ARCI. I limited my study to words written in Arabic. 97% of all the publications found in ARCI have the title written in Arabic and 81% have also an abstract in Arabic. Titles, abstracts and keywords written in Roman script languages are not analysed in this sub-section. It is relatively straightforward to interpret the topics generated by the LDA model. The results are useful to understand the topical structure of ARCI by highlighting the main topics covered in the Arabic scientific literature indexed in ARCI.

**Table 4. 7 topics found in ARCI (2015-2020)**

| *Topic* | *Arabic Tokens* | *Tokens translated into English* |
|---|---|---|
| Islam | محمد ، نبي ، دين ، الله ، علماء | Muhammad, prophet, religion, Allah, scholars |
| Education | تدريب ، رياضيات ، تعلم ، برنامج تعليمي ، طالب | training, mathematics, learning, educational program, student |
| Law | القانون ، الجريمة ، حقوق الإنسان ، القانون الإسلامي ، القانون الدولي | law, crime, human right, Islamic law, international law |
| Finance | بنك إسلامي ، حوكمة ، تأمين ، تمويل ، صكوك | Islamic bank, governance, insurance, finance, sukuk |
| Economics | النمو الاقتصادي ، البطالة ، الاستثمار ، التضخم ، سعر الصرف | economic growth, unemployment, investment, inflation, exchange rate |
| Islamic studies | تفسير ، القرآن الكريم ، الترجمة ، الحديث ، السنة | interpretation, holy Quran, translation, hadith, sunnah |

These topics are not described in details in this study since a certain level of expertise is required to perform such analysis. As mentioned earlier, ARCI provides also some valuable information in English. For instance, 84% of the records found in ARCI have also an abstract in English which can be used to conduct a topic analysis by focusing on the content written in English as well.

### *4.6. Term map*

When applying the LDA model on a corpus, it is assumed one document can address multiple topics. As shown in Table 5, this is helpful to have a precise understanding of the topical structure of a large corpus. However, it does not map the relationships between topics. The purpose of building a so-called term map of the publications in ARCI is to further clarify their contents. I used VOSviewer (van Eck & Waltman, 2010) to create such map.

Titles, abstracts as well as author keywords have been combined into a single string which has been used by the text mining algorithms of VOSviewer. I have limited this analysis to terms which occur at least 15 times. 7,317 terms have satisfied this threshold out of the 259,941 terms found. For each of the 7,317 terms, relevance scores are calculated based on co-occurrence links by VOSviewer. Based on this score, the most relevant terms are shown. VOSviewer offers a default choice of 60% of all the terms. Figure 10 shows the co-occurrence network for the 60% or 4,390 most relevant terms, indicating for each pair of terms the number of papers in which these terms appear together.

The clustering is useful in delineating the topics covered as well as highlighting the relatedness between them. The horizontal and vertical axes have no meaning. The size of a term reflects the number of records in which this specific term is mentioned. The proximity of two terms is an indicator of how these terms are related based on the number of co-occurrences. In general, groups of terms closely located together can be interpreted as topics. Figure 10 displays the term map highlighting the main topics in ARCI (2015-2020). For readability purpose, labels are shown only for selected terms to avoid overlapping labels. The map can also be explored interactively online (https://tinyurl.com/2ol3qxwk) and the labels of the less visible terms phrases can be seen by zooming in on specific map areas.

The terms have been clustered into 7 clusters with different colours. The term map shown in Figure 9 indicates some clear distinctions between research areas. The map confirms a broad coverage of scientific literature as shown previously in the topic analysis. These distinctions are not only visible in the structure in terms of proximity between terms but also in terms of colours. Within an area of the map, terms are usually coloured in a consistent way. For example, the lower left parts in blue and teal includes research areas closely related to Economics and Finance. These clusters include terms like *economic growth, monetary policy, inflation, islamic bank,* and *corporate governance.* In the upper part of the map, purple corresponds to research areas related to Religion and Literature such as *holy quran, translation, interpretation, phonology, rhythm, rhetoric,* and *textual approach*. Terms corresponding to the field of Law tend to be located mainly in the yellow part of the map with terms such as *law, Islamic jurisprudence, contract(s), arbitration, protection* and *justice*. The Religion and History cluster in red is closely related to the Literature and Law parts. It includes terms such as *Allah, religion, peace, prophet, Muhammad (peace be upon him), sunnah* and *Andalusia, century, revolution*. Finally, the lower right parts in green and orange correspond respectively to Education and Sports in general with some distinction on several aspects such as *training program, academic achievement, disorder, psychological empowerment* (green) and *player, physical ability, football* and *physical fitness* (orange).

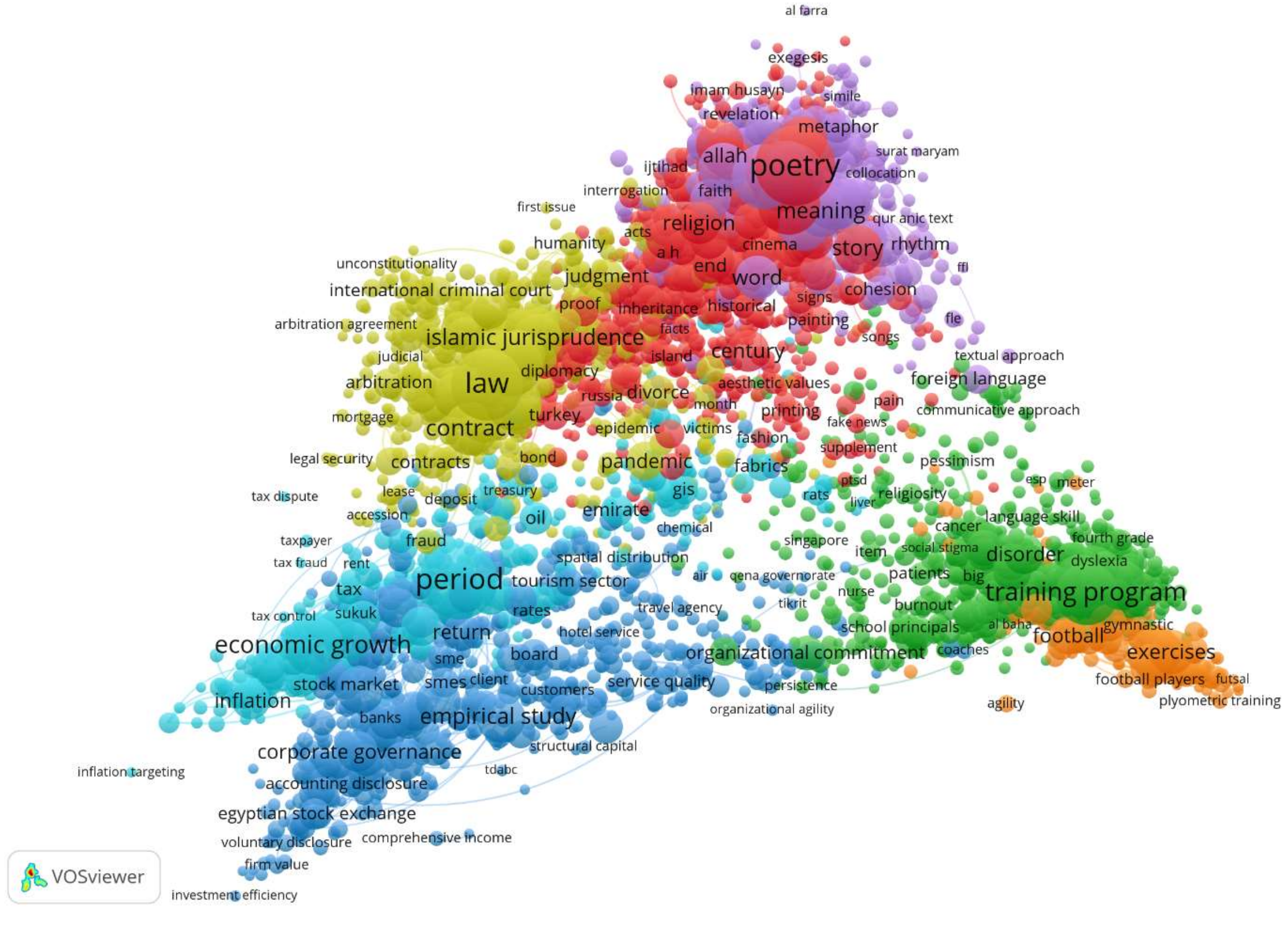

**Fig. 10 Term map highlighting the main topics in ARCI (2015-2020)**

There is a clear heterogeneity in terms of topics covered in ARCI. One should remember some of the terms can of course be used in various contexts. For more information on each cluster, Table 5 lists the 10 most occurring terms in each cluster.

**Table 5. The 10 most occurring terms in each cluster of ARCI (2015-2020)**

| **Red cluster** | # | **Purple cluster** | # | **Yellow cluster** | # |
|---|---|---|---|---|---|
| poetry | 2,488 | interpretation | 1,460 | law | 1,882 |
| discourse | 1,318 | holy Quran | 1,340 | right | 1,390 |
| novel | 1,254 | translation | 1,096 | protection | 1,016 |
| criticism | 987 | text | 1,056 | crime | 907 |
| story | 702 | Quran | 898 | contract | 887 |
| poem | 671 | meaning | 772 | Islamic jurisprudence | 767 |
| Muhammad | 670 | speech | 763 | Algerian legislation | 567 |
| prophet | 522 | hadith | 746 | human right | 547 |
| religion | 499 | significance | 707 | provisions | 546 |
| narration | 478 | grammar | 689 | laws | 503 |
| **Blue cluster** | # | **Teal cluster** | # | **Green cluster** | # |
| empirical study | 711 | period | 2,307 | training program | 1,338 |
| Islamic bank | 660 | economic growth | 1,162 | academic achievement | 935 |
| small | 575 | economic development | 496 | mathematics | 739 |
| corporate governance | 468 | unemployment | 414 | disorder | 456 |
| disclosure | 456 | foreign direct investment | 401 | counselling program | 408 |
| medium enterprise | 419 | inflation | 384 | learning disability | 376 |
| financing | 412 | Algerian economy | 381 | educational program | 355 |
| financial performance | 411 | export | 323 | self esteem | 337 |
| return | 364 | exchange rate | 308 | adolescents | 326 |
| commercial bank | 307 | monetary policy | 289 | student teacher | 319 |
| | | **Orange cluster** | # | | |
| | | exercises | 569 | | |
| | | player | 518 | | |
| | | exercise | 370 | | |
| | | football | 364 | | |
| | | accuracy | 333 | | |
| | | speed | 329 | | |
| | | volleyball | 320 | | |
| | | basketball | 269 | | |
| | | strength | 244 | | |
| | | handball | 201 | | |

### *4.7. Open access*

The last few years have seen the development of several open access (OA) options (Bosman & Kramer, 2018; Lewis, 2012). Open access (OA), and other new technological opportunities such as electronic publishing or open repositories, have changed the scholarly publishing landscape; one effect has been increased accessibility of research output, such as publications. Several scholars have studied the advantage of OA in terms of readership as well as citation impact (Basson et al., 2020; Cintra et al., 2018; Morillo, 2020; Piwowar et al., 2018; Riera & Aibar, 2013; Tang et al., 2017; Torres-Salinas et al., 2019; Young & Brandes, 2020). One of the key issues of recent OA developments has been to understand to which extent the current scientific literature is already published in open access and how that share is evolving in relation to the total growth of the scientific literature. Thus, from research evaluation and management perspectives, it is critical to understand how open access is adopted by the regional research community.

Since 2014, Web of Science has provided information to identify publications from OA journals. The OA status is provided across the Web of Science platform in partnership with the not-for-profit organization *Our Research.* The different types of OA are described as follows:

- *DOAJ Gold:* Journal articles from the Directory of Open Access Journals (DOAJ). To be listed on the DOAJ, each article in these journals needs to have a licence that complies with the Budapest Open Access Initiative.
- *Gold Hybrid:* Other Gold open access papers that are not published in journals on the DOAJ's list but that are identified by *Our Research* as having a Creative Commons (CC) licence. These papers are primarily published in hybrid journals.
- *Free to read:* These articles' licencing is either ambiguous or Our Research has identified them as non-CC licence articles. These are public access or free-to-read articles that can be found on a publisher's website.
- *Green Published:* Final published versions of publications hosted in an institutional or subject-based repository.
- *Green Submitted:* Version of a manuscript that has been submitted and is available in an institutional or topical repository.
- *Green Accepted:* Accepted manuscript hosted in a repository. The final, peer-reviewed content might not have gone through copy-editing or typesetting by the publisher.
- *Non-Open Access:* Publications that do not have an open access status.

I use this information to analyse the access type of records indexed in ARCI. The statistics for various OA types and non-OA records in ARCI are presented in Table 6.

**Table 6. Number and share of records by Open-Access type in ARCI (2015-2020)**

| *Open access type* | *Records* | *Share (%)* |
|---|---|---|
| Non Open Access | 84,853 | 61.4 |
| All Open Access | 43,230 | 31.3 |
| *Free to Read (Bronze)* | 29,927 | 21.6 |
| *DOAJ Gold* | 11,386 | 8.2 |
| *Gold Hybrid* | 1,067 | 0.8 |
| *Green Published* | 711 | 0.5 |
| *Green Submitted* | 426 | 0.3 |
| *Green Accepted* | 5 | 0.0 |
| Unknown (no DOI) | 10,200 | 7.4 |

Close to 31% of papers indexed in ARCI and published between 2015 and 2020 are openly accessible. This is below the average share of 36% of OA documents in the Web of Science Core Collection for the same period. We notice the various OA types have different shares in ARCI. *Free to read* or *Bronze* is the main OA type with 29,927 papers representing about 21% of ARCI. *DOAJ Gold* has the second highest OA share (8.2%) in this database with 11,386 papers published with this OA type. Also, it is worth noting that the OA type is unknown for 10,200 papers in ARCI (7.4%) which do not have a DOI.

Figure 11 shows the trends of OA shares in ARCI and WOS between 2015 and 2020. Due to their relatively low shares, Gold-Hybrid, Green published, Green Submitted and *Green Accepted* were excluded from this analysis. ARCI shows a lower share of all OA documents compared to WOS, but ARCI presents a similar uptrend. In terms of the share of *Gold* OA in ARCI, although it has been increasing since 2017, it represents about half the share of *Gold OA* in WOS. Lastly, the share of *Bronze* OA in ARCI has been stable since 2015 and is about 3 times higher than in WOS.

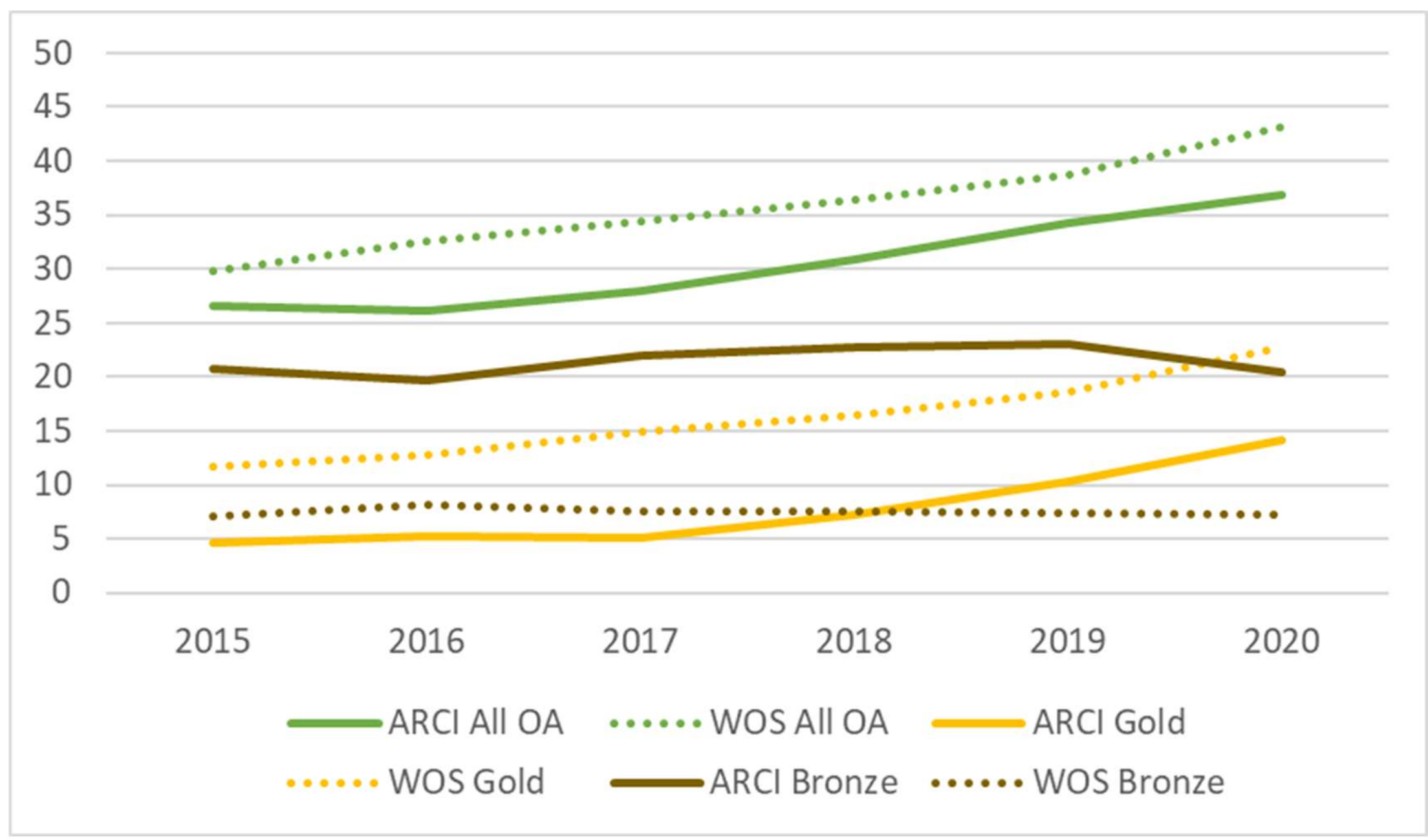

**Fig. 11 Trends of shares of records by OA type in ARCI and in WOS (2015-2020)**

Such analysis is useful in terms of research management and OA organisation. ARCI provides insights on OA practices in the Arab region. While the original goal of open access was to make scientific findings accessible to everyone, one development was to shift the expenses of Article Processing Charges (APC) to the scientific authors. Although scientific papers published in OA can be accessed by everyone, research cannot always be published under the APC-based OA model. Indeed, since Arab League nations are mainly developing or emerging countries, the overall lower share of OA content in ARCI compared to WOS, might be due to high APCs that researchers might not be able to afford without proper funding mechanisms. However, it is also important to remind other ways in which OA can be realized such as Diamond, Bronze and Green OA which do not have APCs.

Currently, many governmental institutions fund the publication of journals and/or support the cost of journal subscriptions with public funds. They also have now to cover the funding of APC. The Global Open Access Portal (GOAP) presented a snapshot of the status of Open Access (OA) to scientific information worldwide. As identified in the GOAP, the Arab States face challenges but also opportunities (Unesco, 2016). Low level of awareness of the OA potential for researchers, publishers and policy makers tops the list of challenges. Lack of policy regulation, research funders' OA mandates and resources to manage OA projects also contribute to the low OA penetration in the Arab world. Nevertheless, several projects and initiatives have been undertaken already to promote OA in the Arab region. 70 experts and policy specialists from several Arab Countries met in September 2015 to develop strategies to implement open access to scientific information and research in the Arab countries (Unesco, 2015). The Directory of Free Arab Journals (DFAJ), the first Arab directory of Open Access Journals which provides access to journals published by 172 publishers from 17 Arab countries, is also an example of such initiative.

## 5. Discussion

The main objective of this study was to examine the structure of ARCI, the Arabic Citation Index. As of June 2021, 613 Arabic journals were indexed in ARCI. This indexation brings several benefits to the scientific community. This new index will improve the visibility of Arabic journals by making them more accessible. All journals indexed in ARCI need to meet selection criteria and essential publication metadata are provided. Such a database could greatly enhance scholarly literature search. As a result, this will also help researchers to identify critical and influential research published in Arabic.

Research evaluation often implies the bibliometric analysis of research output (Wilsdon et al., 2015; Wouters et al., 2015). Bibliometrics analysis plays an important role in research policies in many countries. Such policies involve the usage of bibliometric databases to evaluate research at various levels such as national, institutional or author level. Indicators based on citation indices are now widely used in academic assessments (Bornmann & Haunschild, 2018; Campbell et al., 2010; Derrick & Pavone, 2013; Hicks & Melkers, 2013). ARCI could provide useful bibliometric data sources to research managers for science assessment and research analysis. This would be helpful to *identify and reward excellence in locally relevant research* (Hicks et al., 2015). ARCI is also likely to attract attention from publishers and funders.

Now, I discuss in detail the main findings identified in this analysis. The main objective was to provide a brief profile of ARCI. This study reveals that ARCI contains mainly journals in the Arts & Humanities and Social Sciences categories. It is important to keep in mind the well documented limitation on subject delineation where I used research areas of journals as proxies of categories to characterise the subject coverage and to understand the level of contribution and specialisation of each country. Egypt, Algeria, Iraq and Saudi Arabia contribute to most research categories.

As per the analysis of the publications covering the 2015-2020 period, ARCI indexes content from 19 of the 22 Arab League countries, with more than 60% of the journals indexed in ARCI being published in Egypt and Algeria, and more than 22% published in Iraq, Jordan and Saudi Arabia. As mentioned previously, several Ministries of Higher Education and Research have set up initiatives to improve the visibility of local journals. Such initiatives include the development of national journal platforms with a standardisation of the journals' meta data, as well as the delivery of workshops with journal editors to improve their publishing standards. Since ARCI is still new and under development, it will be interesting to track its coverage growth over time.

It is worth reminding that the country of publisher of the journal is considered for its indexation in ARCI and not only the language of publication. Thus, ARCI does not include yet the journals published in Arabic in countries which are not members of the Arab League. With no surprise, the analysis of research output by country also reveals a concentration of publications by authors affiliated to institutions in Arab League nations. However, some cases standout such as Iran, Malaysia, the United States of America, France, Turkey or the United Kingdom which are among the top 25 contributing countries to ARCI in terms of authors' countries. As of now, most of the content found in ARCI is composed of articles (98.9% of the database). Since the Humanities tend to traditionally rely on book chapters and books, it will be interesting to analyse the evolution of the coverage by document type. Unsurprisingly, ARCI has a great share of papers published in Arabic (about 92% of the database). However, English and French are two other languages well represented in ARCI. Other languages suggest research published in ARCI journals may also tackle regional issues of interest with neighbour regions such as Europe and Asia.

Analysing the authorship structure in ARCI provides a better understanding of the specific dynamics involved in the production of scientific knowledge and the development of research policy. With the most common type of authorship in ARCI being single, there is a strong preference for single work, which is not surprising considering single authorship as a common practice in humanities and social sciences. This might also suggest a relatively low level of international or regional collaborations between researchers from the Arab League nations. Fields such as Cultural Studies, Quranic Studies, Poetry, Hadith, Islamic Creed and Social work all show a share of single authorship publications higher than 90%. On the other hand, multi authorship is more frequent in other disciplines such as Geography, Special Education, Management, and Economics, which suggests that those areas exhibit a more collaborative aspect.

The topic analysis as well as the term map are helpful to better understand the underlying structure of ARCI. Such techniques provide a great overview of the topics covered in this database. Overall, the clusters found with VOSviewer seem to be closely related and show a broad coverage of ARCI. ARCI also offers the possibility to analyse the corpus in Arabic. The terms found provide useful information about the topics of regional relevance.

Around 31% of the content indexed in ARCI is openly accessible, which is below the share of open access publications indexed in WOS (36%) in the same period. The Open Access information available in ARCI is particularly useful to better share scientific knowledge as well as to track the adoption of local OA mandates by research managers. The insights provided by ARCI can help agencies and academic institutions in the development of policies of strategic planning and for APC funding. As a recommendation, there should be a better awareness of the existing OA publication model among research institutions and researchers. The demand for funding and funding policies for the publication of papers should consequently increase in the coming years. Indeed, national governments might be able to stimulate their publishing capabilities (Moed et al., 2021). They may also establish criteria or formulas for academic institution funding as well as staff evaluation for recruiting and advancement. To monitor the success of their financing strategies, research funders can also use data from journals and other sources. In the context of research management and research evaluation, information available in ARCI can be used to assess and inform research activities and performance of research stakeholders at various levels (individuals, groups, institutions, or national systems). Last but not least, individual researchers can use the literature indexed in ARCI for their daily scholarly activities.

In conclusion, this paper offers a profile of the newest citation index in the WOS. This paper contributes also to the literature on regional citation indices (Huang et al., 2017; Jin & Wang, 1999; Leydesdorff & Jin, 2005; Moskaleva et al., 2018; Pajic, 2015; Seol & Park, 2008; Velez-Cuartas et al., 2016). One common purpose of such regional databases is to provide more visibility to local journals and research published in other languages than English. As of January 2020, Arabic was the fourth most popular language online with 5.2% of worldwide internet users, following English (25.9%), Chinese (19.4%) and Spanish (7.9%) and it is also the fastest-growing language on the internet in terms of number of Internet users as of 2021 (Statista, 2022). With such observation, one can predict the potential increase of scientific content published in Arabic as well. Thus, ARCI is likely to have positive effects on regional research discovery as well as research management and research evaluation in the Arab region. Indeed, multidisciplinary databases, like the WOS, only provide a partial picture of research publishing activities, particularly for non-English scientific publications. These effects are still too early to see, but ARCI sets strong foundations for a more inclusive research evaluation framework in the MENA region or more specifically in Arab league nations. Future research may seek to propose detailed mappings of ARCI to better understand its structure and impact. Finally, it will also be interesting to track its expansion and evolution by using dynamic topic models to study the time evolution of topics by using the text available in English as well as Arabic.

## Acknowledgments

A preliminary version of this paper has been presented at the ISSI 2021 conference (El-Ouahi, 2021). I want to express my gratitude to Ludo Waltman and Thomas Franssen for their insightful comments on a previous version of the manuscript. I am also grateful for the feedback provided by two reviewers.

## Competing Interests

The author is an employee of Clarivate Analytics, the provider of the Web of Science and the Arabic Citation Index.

## Data Availability

The research presented in this paper uses Web of Science data made available by Clarivate. The author is not allowed to share this data.